\documentclass[epj,a4paper,nofootinbib,nobibnotes,twocolumn,groupedaddress,floatfix]{svjour}
\usepackage{graphics}
\usepackage{epsfig}
\usepackage{graphicx}
\usepackage[square, sort&compress]{natbib}
\usepackage{amsmath}
\usepackage{har2nat}
\usepackage{color}%
\usepackage{dcolumn}
\usepackage[utf8]{inputenc}
\usepackage[english]{babel}
\usepackage{verbatim}   
\usepackage{amssymb}

\begin{document}

\title{Setup commissioning for an improved measurement of the  D(p,$\gamma$)$^3$He cross section at Big Bang Nucleosynthesis energies}

\titlerunning{Setup commissioning for improved D(p,$\gamma$)$^3$He cross sections}

\authorrunning{V. Mossa {\it et al.} (LUNA collaboration)}

\author{V.~Mossa \inst{1} \and 
K.~St\"ockel  \inst{2,3} \and 
F.~Cavanna \inst{4} \and 
F.~Ferraro \inst{4,5} \and 
M.~Aliotta \inst{6} \and 
F.~Barile \inst{1} \and 
D.~Bemmerer \inst{2} \and 
A.~Best \inst{7} \and 
A.~Boeltzig \inst{8,9} \and 
C.~Broggini \inst{10} \and 
C.G.~Bruno \inst{6} \and 
A.~Caciolli \inst{10,11} \and 
L.~Csedreki \inst{8,9} \and 
T.~Chillery \inst{6} \and 
G.F.~Ciani \inst{8} \and 
P.~Corvisiero \inst{4,5} \and 
T.~Davinson \inst{6} \and 
R.~Depalo \inst{10} \and 
A.~Di Leva \inst{7} \and 
Z.~Elekes \inst{12} \and 
E.M.~Fiore \inst{1,13} \and 
A.~Formicola \inst{9} \and 
Zs.~F\"ul\"op \inst{12} \and 
G.~Gervino \inst{14} \and 
A.~Guglielmetti \inst{15} \and 
C.~Gustavino \inst{16}\thanks{Corresponding author: carlo.gustavino@roma1.infn.it} \and 
G.~Gy\"urky \inst{12} \and 
G.~Imbriani \inst{7} \and
M.~Junker \inst{9} \and 
I.~Kochanek \inst{9} \and 
M.~Lugaro \inst{17} \and 
L.E.~Marcucci \inst{18} \and 
P.~Marigo \inst{10,11} \and 
E.~Masha \inst{15} \and 
R.~Menegazzo \inst{10} \and 
F.R.~Pantaleo  \inst{1,19} \and 
V.~Paticchio \inst{1} \and 
R.~Perrino  \inst{1}\thanks{Permanent address: INFN Sezione di Lecce, Lecce, Italy} \and 
D.~Piatti \inst{10} \and 
P.~Prati \inst{4,5} \and 
L.~Schiavulli \inst{1,13} \and 
O.~Straniero \inst{20} \and 
T.~Sz\"ucs  \inst{2} \and 
M.~P.~Tak\'acs  \inst{2,3} \and 
D.~Trezzi \inst{15} \and 
S.~Zavatarelli \inst{4}\thanks{Corresponding author: sandra.zavatarelli@ge.infn.it} \and 
G.~Zorzi \inst{15}
\newline (LUNA collaboration)
}

\institute{ %
INFN, Sezione di Bari, via E. Orabona 4, 70125 Bari, Italy \and  
Helmholtz-Zentrum Dresden-Rossendorf, Bautzner Landstraße 400, 01328 Dresden, Germany \and  
Technische Universit\"at Dresden, Zellescher Weg 19, 01069 Dresden, Germany \and  
INFN, Sezione di Genova, Via Dodecaneso 33, 16146 Genova, Italy \and  
Universit\`a degli Studi di Genova, Via Dodecaneso 33, 16146 Genova, Italy \and	
SUPA, School of Physics and Astrophysics, University of Edinburgh, Peter Guthrie Tait Road, EH9 3FD Edinburgh, United Kingdom \and	
Universit\`a degli Studi di Napoli ``Federico II'', and INFN, Sezione di Napoli, Via Cintia 21, 80126 Napoli, Italy \and	
Gran Sasso Science Institute, Viale F. Crispi 7, 67100 L’Aquila, Italy \and  
INFN, Laboratori Nazionali del Gran Sasso (LNGS), Via G. Acitelli 22, 67100 Assergi, Italy \and 
INFN, Sezione di Padova, Via F. Marzolo 8, 35131 Padova, Italy \and  
Universit\`a degli Studi di Padova, Via F. Marzolo 8, 35131 Padova, Italy \and 
Institute for Nuclear Research (Atomki), PO Box 51, 4001 Debrecen, Hungary \and  
Universit\`a degli Studi di Bari, Dipartimento Interateneo di Fisica, Via G. Amendola 173, 70126 Bari, Italy \and  
Universit\`a degli Studi di Torino and INFN, Sezione di Torino, Via P. Giuria 1, 10125 Torino, Italy \and  
Universit\`a degli Studi di Milano and INFN, Sezione di Milano, Via G. Celoria 16, 20133 Milano, Italy \and 
INFN, Sezione di Roma, Piazzale A. Moro 2, 00185 Roma, Italy \and  
Konkoly Observatory, Research Centre for Astronomy and Earth Sciences, MTA Centre for Excellence, Konkoly Thege Miklos 15-17, H-1121 Budapest, Hungary \and 
Universit\`a degli Studi di Pisa and INFN, Sezione di Pisa, Largo Bruno Pontecorvo, 56127 Pisa, Italy \and  
Politecnico di Bari, via Amendola, 126 b, 70126 Bari, Italy \and  
INAF Osservatorio Astronomico d'Abruzzo, Via Mentore Maggini, 64100 Teramo, Italy 
}

\date{\today}

\abstract{Among the reactions involved in the production and destruction of deuterium during Big Bang Nucleosynthesis, the deuterium-burning D(p,$\gamma$)$^3$He reaction has the largest uncertainty and limits the precision of theoretical estimates of primordial deuterium abundance.
Here we report the results of a careful commissioning of the experimental setup used to measure the cross-section of the D(p,$\gamma$)$^3$He reaction at the Laboratory for Underground Nuclear Astrophysics of the Gran Sasso Laboratory (Italy). 
The commissioning was aimed at minimising all sources of systematic uncertainty in the measured cross sections. 
The overall systematic error achieved ($< 3\%$) will enable improved predictions of BBN deuterium abundance.}

\maketitle

\section{Primordial deuterium abundance and the $\mathrm{\mathbf{D(p,\gamma)^{3}He}}$ reaction}
Big Bang Nucleosynthesis (BBN) occurs during the first minutes of cosmological time in a rapidly expanding hot and dense Universe, where a fraction of protons and nearly all free neutrons end up bound in  $^4$He, while D, $^3$H, $^3$He, $^6$Li, $^7$Li and $^7$Be nuclei form in trace quantities \cite{Serpico2004}. 
As the nucleosynthesis of these primordial elements is sensitive to the physics of the early universe, their abundance can be used to test the standard model of modern cosmology and particle physics. 
The primordial abundance of deuterium, in particular, can provide stringent constraints on the baryon density and the number of relativistic particles in the early Universe \cite{Pitrou18-PR,DiValentino14-PRD}.

Recent astronomical observations of primordial deuterium abundance have reported values with a 1\% uncertainty \cite{Cooke18-ApJ}.
By contrast, theoretical BBN calculations are still hindered by relatively large uncertainties 
in the D(p,$\gamma$)$^3$He reaction cross section, which remains at present the least well-known of all reactions involved in the nucleosynthesis of primordial deuterium \cite{Pitrou18-PR,DiValentino14-PRD,Coc15-PRD}.
In order for BBN predictions to achieve the same precision as observations, the D(p,$\gamma$)$^3$He cross section must be determined with a systematic uncertainty better than 3\% at energies relevant to BBN (see for example, Refs. \cite{Coc15-PRD,Pitrou18-PR}). 

Several data sets on the D(p,$\gamma$)$^3$He reaction cross section, or equivalently its $S(E)$-factor\footnote{The astrophysical $S(E)$ factor is defined as \cite{Rolfs88-Book}: $S(E)=E\sigma(E)\exp{(2\pi\eta)}$, where $E$ is the energy of interaction in the centre of mass system, $\sigma(E)$ is the energy dependent cross-section, and $\eta$ is the Sommerfeld parameter $\eta(E)={\rm Z_1 Z_2}\alpha (\mu c^2/2E)^{1/2}$ (with Z$_i$ atomic numbers of the interacting particles, $\alpha$ fine structure constant, $\mu$ reduced mass, and $c$ speed of light).}, are available in the literature.
In the low-energy range ($E_{\rm cm} \simeq 2-20$~keV), mostly relevant to hydrogen burning in the Sun and in protostars, cross sections were obtained with a systematic error of at most 5.3\% \cite{Casella02-NPA} using the $50$~kV accelerator (now in disuse) of the 
Laboratory for Underground Nuclear Astrophysics (LUNA) at the Laboratori Nazionali del Gran Sasso (LNGS) in Italy  \cite{Broggini18-PPNP,Cavanna18-IJMPA}.
At higher energies ($E_{\rm cm} \simeq 30-700$~keV), available data sets are affected by systematic errors of 9\% or higher \cite{Ma97-PRC,Schmid97-PRC,Griffiths63-CJP,Tisma19-EPJA}. 
The situation is further compounded by the fact that a recent \emph{ab initio} calculation \cite{Marcucci16-PRL} disagrees at the 20-30\% level with both the $S$-factor of Ma {\em et al.} \cite{Ma97-PRC} and a best fit  \cite{Adelberger11-RMP} to selected data \cite{Griffiths62-CJP,Schmid96-PRL,Ma97-PRC,Casella02-NPA}, widely used in BBN calculations.

Here, we report the results of setup commissioning measurements aimed at minimising all sources of systematic uncertainties in the measurement of the D(p,$\gamma$)$^3$He cross section. We have exploited the full dynamic range of the 400~kV accelerator, corresponding to $E_{\rm cm} \simeq 30-300$~keV and thus covering a broad energy region around $E_{\rm cm} \simeq 130$~keV where, according to Nollett {\em et al.} \cite{Nollett00-PRD}, the predicted abundance of primordial deuterium is most sensitive to the D(p,$\gamma$)$^3$He reaction.

The paper is organised as follows: first, we describe the experimental approach and setup used at LUNA for the measurement of the D(p,$\gamma$)$^3$He cross section (sect. \ref{sec:setup}); then, we report on high-precision measurements of target density profile, beam current, detection efficiency, and angular distribution effects
(sections \ref{sec:gt}-\ref{sec:PSA}). 
The results obtained are discussed and summarised in sect. \ref{sec:errors}.

\section{Experimental approach and setup}
\label{sec:setup}

The D(p,$\gamma$)$^3$He reaction ($Q$-value = 5.5~MeV) was studied in direct kinematics using a windowless and extended D$_2$ gas target and detecting the emitted $\gamma$ rays with a high purity germanium detector. 
At LUNA, the $\gamma$-ray background is reduced by more than four orders of magnitude in the region of interest for the D(p,$\gamma$)$^3$He reaction ($E_{\gamma} = 5-6$~MeV) \cite{Broggini18-PPNP}.

For an extended gas target of length $L$, the cross section can be expressed in terms of experimentally measurable quantities as:
\begin{equation}
    \sigma(E)=\frac{N_{\gamma}(E)}{N_p\int_{-L/2}^{L/2}\rho(z)\epsilon(z,E_\gamma)W(z)dz}
    \label{eq:cs}
\end{equation}
where $N_{\gamma}(E)$ is the net number of detected $\gamma$ rays at a given interaction energy $E$, $N_p$ the number of incident protons, $\rho(z)$ the number density of target atoms as a function of interaction position $z$ along the target,
$\epsilon(z,E_\gamma)$ the $\gamma$-ray photo-peak detection efficiency and $W(z)$ a term accounting for the angular distribution of the emitted gamma rays. 
The integral in the denominator extends over the whole gas target length.
To achieve a 3\% systematic uncertainty on the cross section, each of the quantities in eq. (\ref{eq:cs}) must be determined with an appropriate level of accuracy, as will be explained in the following sections.

\begin{figure*}[h!]
\centering
\includegraphics[width=35pc]{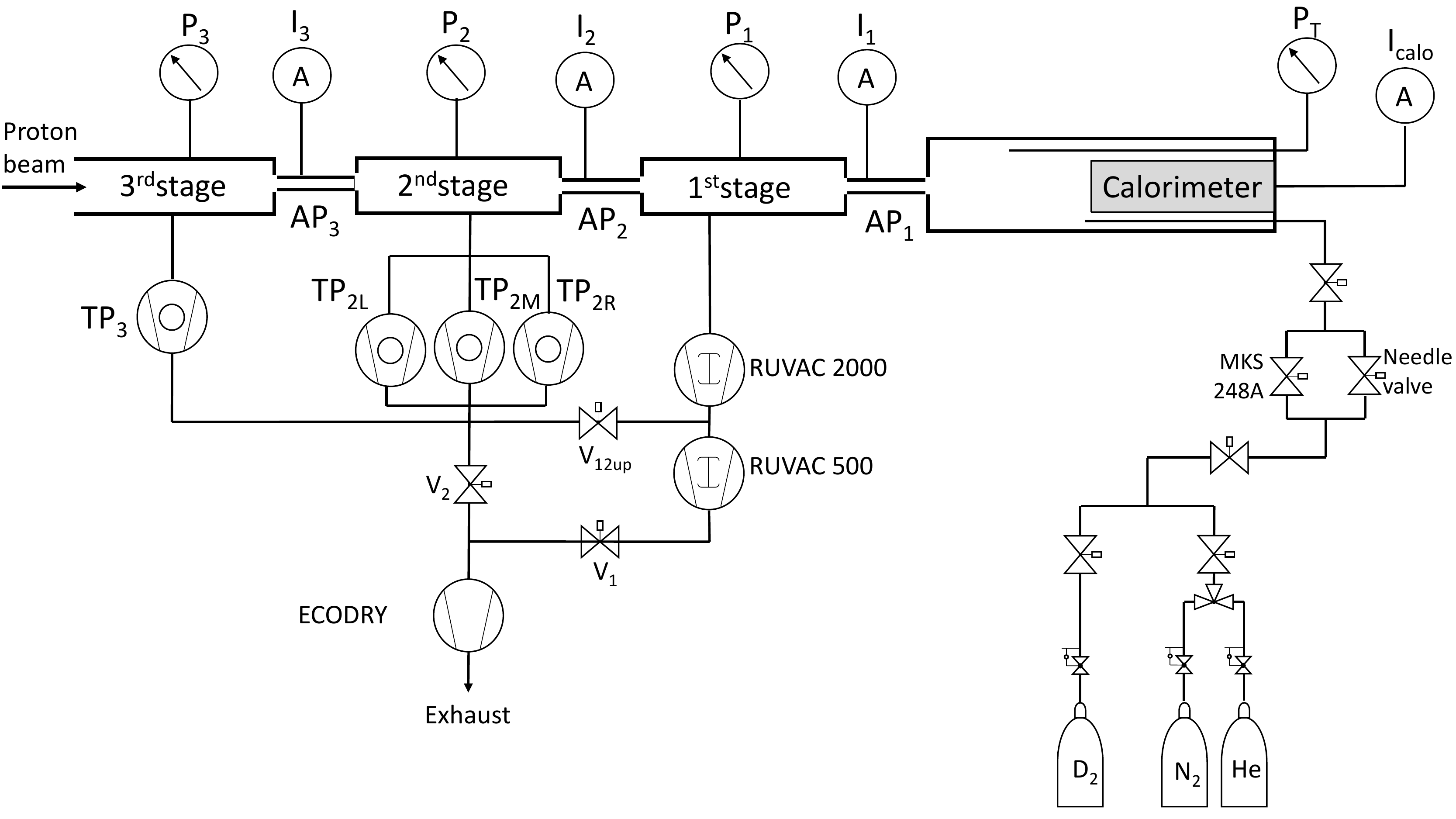}\hspace{2pc}
\caption{Scheme of the LUNA windowless gas target (not to scale). Shown are the three pumping stages and the target chamber with the calorimeter.}
\label{fig:gastarget}
\end{figure*}

The experimental setup used is shown in fig. \ref{fig:gastarget}.
Briefly, the LUNA 400~kV accelerator was used to provide a proton beam at energies $E_{\rm p} = 50-400$~keV (with a systematic error of 0.3~keV \cite{Formicola03-NIMA}) and typical intensities of 200~$\mu$A.
The beam enters the target chamber after passing three differential pumping stages connected through a series of apertures of 25, 15, and 7~mm diameter (hereafter AP3, AP2 and AP1, see fig. \ref{fig:gastarget}) \cite{Cavanna14-EPJA}. Collimators AP3 and AP2 are upstream at a distance of 69 and 51~cm, respectively, from the entrance of the target chamber. Collimator AP1 is 40 mm long and is mounted at the entrance of the target chamber. This aperture is designed to be sufficiently long and narrow to enable a pressure drop of a factor of 20 between the target chamber and the first pumping stage.  

Deuterium gas of 99.99\% isotopic purity (as certified by the vendor, Rivoira \cite{rivoira}) was maintained at a pressure of 0.3 mbar inside the target chamber by means of a capacitive pressure sensor (independent of gas type, calibrated to 0.25\% accuracy) controlling the gas inlet via a feedback loop. 
The target chamber, a 330~mm long cylinder with an inner diameter of 56~mm, is made of 3~mm thick AISI 304 stainless steel and is equipped with a series of ports to measure the gas target temperature and pressure along the beam axis (see sect. \ref{sec:gt}). 
The differential pumping system maintains a modest vacuum at the first pumping stage, evacuated by a 2050 m$^3$/h Roots pump. 
The second pumping stage is equipped with three 1500 l/s turbomolecular pumps, and the third pumping stage with a 360 l/s turbo-molecular pump. Typical pressures observed during the experiment, with 0.3 mbar of deuterium in the target chamber, were: 1.5$\times$10$^{-3}$ mbar in the first stage, 2.7$\times$10$^{-6}$ mbar in the second stage, and 2.6$\times$10$^{-6}$ mbar in the third stage. 
Fresh deuterium was flushed without re-circulation to prevent its pollution. Possible  air-leaks contamination (mostly nitrogen) was periodically checked by using the well-known $^{14}$N(p,$\gamma$)$^{15}$O resonance at $E_{\rm r}= 259$~keV \cite{Daigle16-PRC}. 
Specifically, we exploited the relationship between the resonance strength and the effective stopping power (see for example, \cite{Rolfs88-Book}) to determine the ratio of nitrogen to deuterium atoms in the gas. This ratio was found to be always below 0.5\%.

The target chamber contains a cylindrical copper ca\-lo\-ri\-meter (30~cm length and 50~mm diameter) for beam current integration (sect. \ref{sec:current}). Gamma rays from the D(p,$\gamma$)$^3$He reaction were detected with a high purity germanium detector (not shown in the figure, relative efficiency\footnote{The relative efficiency is  defined at 1.33 MeV relative to that of a standard 3”-diameter, 3”-long NaI(Tl) scintillator at 25~cm from the source.} 135\%) located under the target chamber and facing its centre.  

\section{Effective target density measurements}
\label{sec:gt}
\begin{figure}[h!]
\centering
\includegraphics[width=21.5pc]{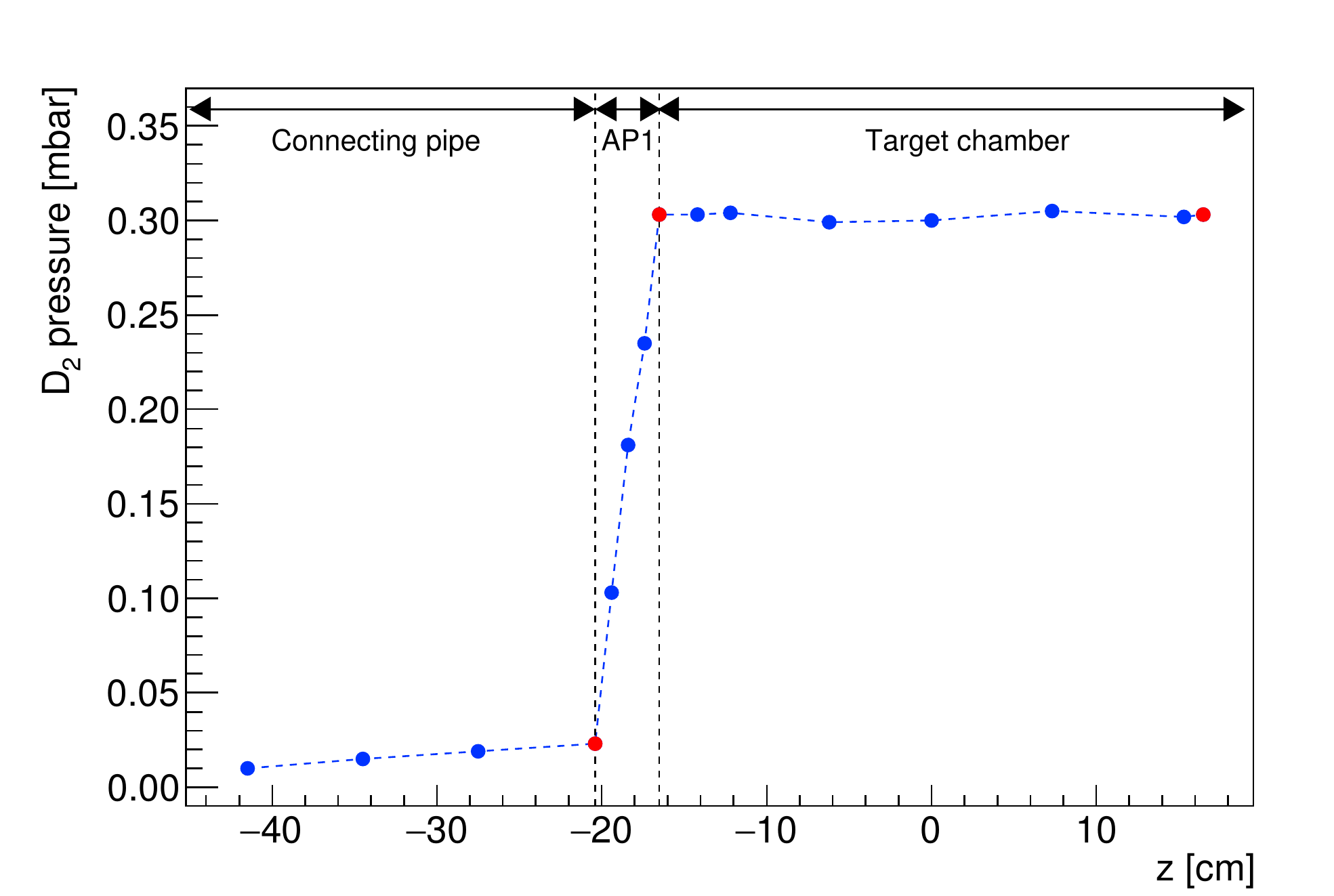}\hspace{2pc}%
\caption{Blue points are the measured pressure profile in deuterium along the beam axis ($z=0$ corresponds to the centre of the target chamber). Red points are the extrapolated pressure values at the collimator interfaces with the connecting pipe and the target chamber and at the calorimeter surface. The lines are drawn to guide the eye.}
\label{fig:pprofile}
\end{figure}

The gas target density $\rho(z)$ was determined by combining independent measurements of pressure and temperature profiles along the beam axis.  Measurements were performed without beam and under the same operating conditions: stable deuterium pressure inside the gas target ($P=0.3$~mbar), water-cooled AP1 collimator, and beam stop temperature maintained at 343~K (70$^o$~C, see sect. \ref{sec:current}).
For the pressure profile, six capacitance gauges (five Pfeiffer CMR 363 and one MKS Baratron 626, typical precision 0.3\%) were used to measure the pressure at 12 different positions: six inside the target chamber, three in the AP1 collimator and three in the pipe connecting the chamber to the first pumping stage. 
As shown in fig. \ref{fig:pprofile}, the gas pressure remains constant within $\pm 0.9\%$ inside the chamber, decreases along the collimator, and vanishes within the connecting pipe. 
Taking into account calibration and instrumental accuracy, a total uncertainty of $\pm 0.9$\% was assigned to the pressure inside the target chamber. 

\begin{figure}[h!]
\centering
\includegraphics[width=22pc]{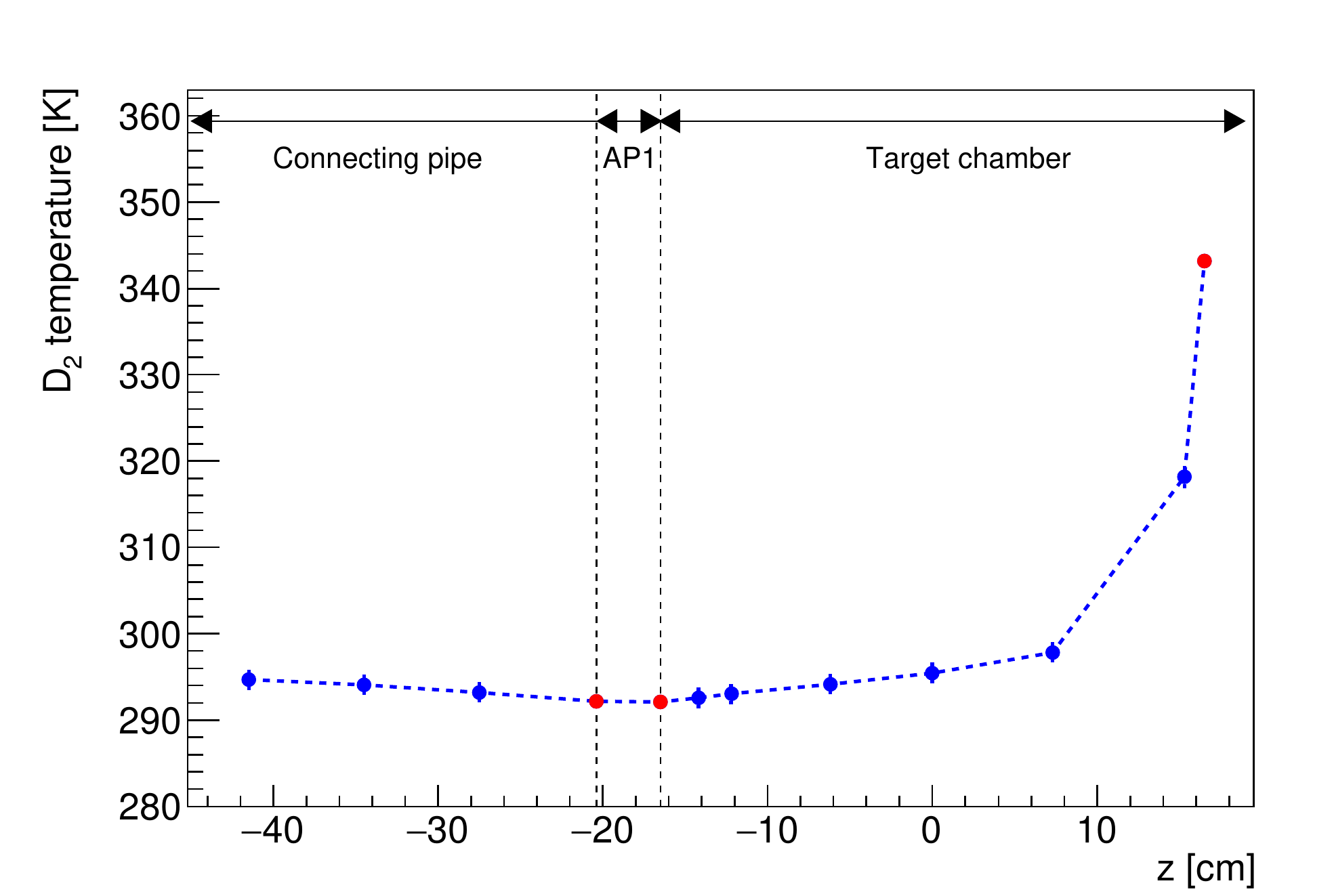}\hspace{2pc}%
\caption{Blue points are the measured temperature profile in deuterium along the beam axis ($z=0$ corresponds to the centre of the target chamber), at a pressure of 0.3~mbar inside the target chamber. Red points are the extrapolated temperature values at the collimator interfaces with the connecting pipe and the target chamber and the calorimeter hot surface temperature. The lines are drawn to guide the eye.}
\label{fig:tprofile}
\end{figure}

The gas temperature was measured at 12 positions (along the beam axis) inside the chamber and the connecting pipe using four PT100 thermistors. The measured profile is shown in fig. \ref{fig:tprofile}: the temperature drops monotonically between the calorimeter beam stop heated to 343~K (see sect. \ref{sec:current}) and the collimator, while it remains constant to within $\pm 0.2$\% inside the connecting pipe at about 294~K (21$^o$~C). 
The overall temperature error was estimated to be about 1~K, corresponding to a relative error of $\pm 0.3$\%.

\begin{figure}[t!]
\includegraphics[width=20pc]{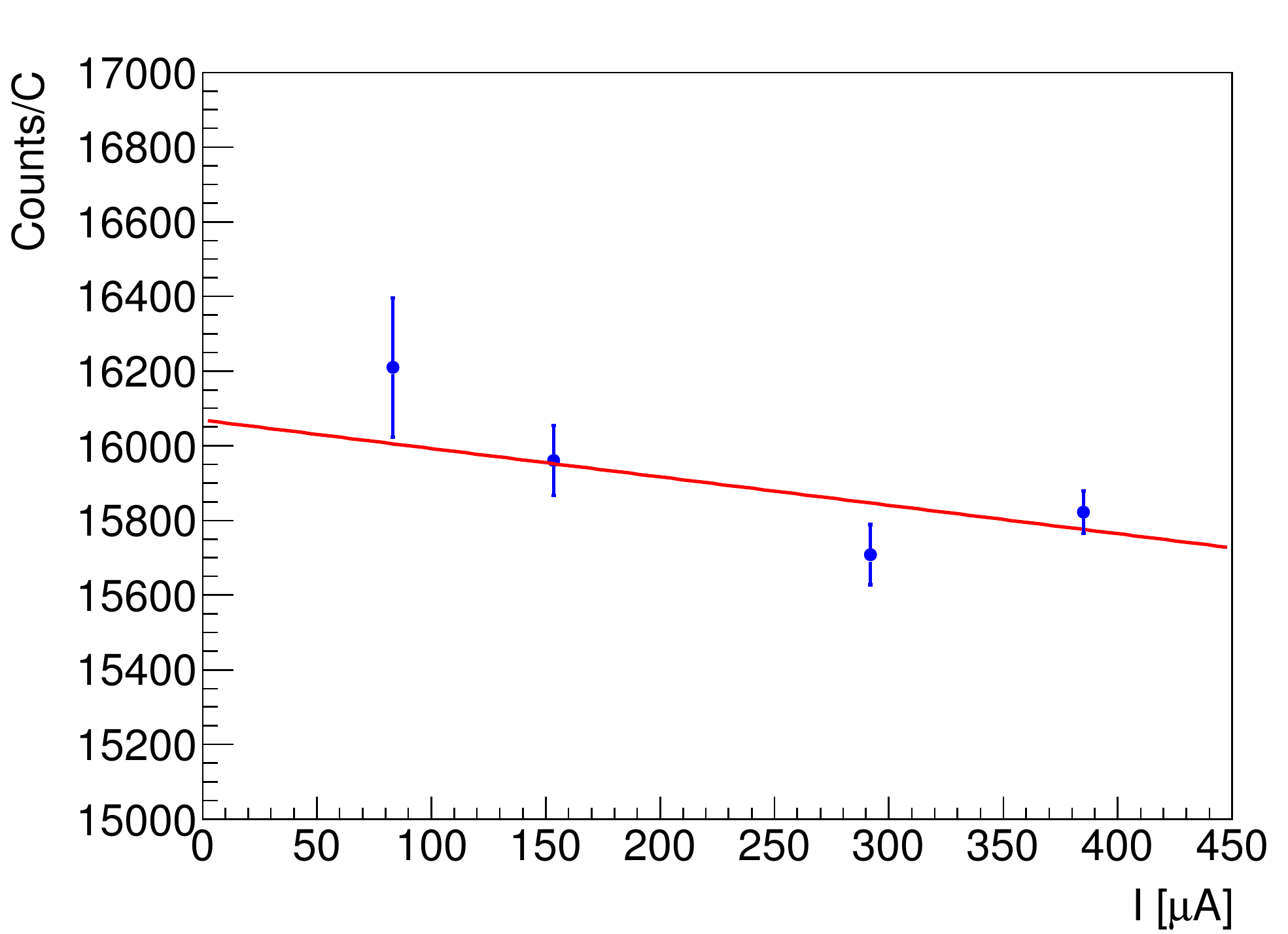}\hspace{2pc}%
\caption{Counts per unit charge vs beam current in a D$_2$ gas target at $P=0.3$~mbar and $E_{\mathrm{beam}}=300$ keV.}
\label{fig:BH}
\end{figure}

It should be noted, however, that during actual measurement runs the gas temperature (and thus the gas density) can be affected by the power dissipated by the beam\footnote{The power dissipated by the beam is calculated as the product of the beam intensity and the beam energy lost in the gas.
For our experiment, we obtain $23.3$~mW/cm ($8.5$~mW/cm) at the lowest (highest) beam energy $E_{\rm p} = 50$~keV ($E_{\rm p} = 395$~keV) used.} in passing through the gas (beam heating effect \cite{Rolfs88-Book}).
The extent of this effect was  quantified experimentally by measuring the D(p,$\gamma$)$^3$He reaction yield for different beam intensities at a fixed beam energy and constant deuterium pressure (0.3~mbar). 
Figure \ref{fig:BH} shows the results obtained for a 300~keV proton beam: a typical current of 200~$\mu$A leads to a count rate, {\em i.e.} target density, reduction of about 1\%, in good agreement with previous findings obtained at LUNA with an alpha beam and a deuterium gas target \cite{Anders13-EPJA, MARTA2006} and also with an analytical expression for beam heating effect in gas targets reported in \cite{osborne84}. We adopted a $\pm 50$\% uncertainty on the 1\% correction factor due to the beam-heating effect. 
The overall resulting uncertainty (1.1\%) on the inferred target density is then obtained in quadrature from the uncertainties in temperature (0.3\%), pressure (0.9\%), and  beam heating effect (0.5\%).

\section{Beam current measurement}
\label{sec:current}

As the proton beam passes through the gas target, protons are partially neutralized and secondary electrons are produced.
These phenomena prevent a reliable electrical reading of the beam current and thus of the number of protons N$_p$ impinging on the target. 
For this reason the beam current was measured with a calorimeter maintained at a constant temperature gradient. 
The calorimeter consists of three parts: a hot side (acting as a beam stop), eight heating resistors, and a cold side. 
The hot side is kept at a temperature of 70~$^{\circ}$C by regulating the current of the heating resistors via a feedback loop in which four thermistors (PT100) are used to monitor the temperature of the beam stop \cite{Casella02-NIMA}. The cold side of the calorimeter is kept at 0~$^{\circ}$C by means of a chiller. 
The beam stop can be heated either by the resistors or by the beam, thus the more power is dissipated by the beam in the calorimeter, the less has to be provided by the resistors to maintain the temperature gradient. Power values and temperatures of the hot and cold sides were continually recorded by a real time controller programmed in LabVIEW.

\begin{figure}[t!]
\includegraphics[width=21pc]{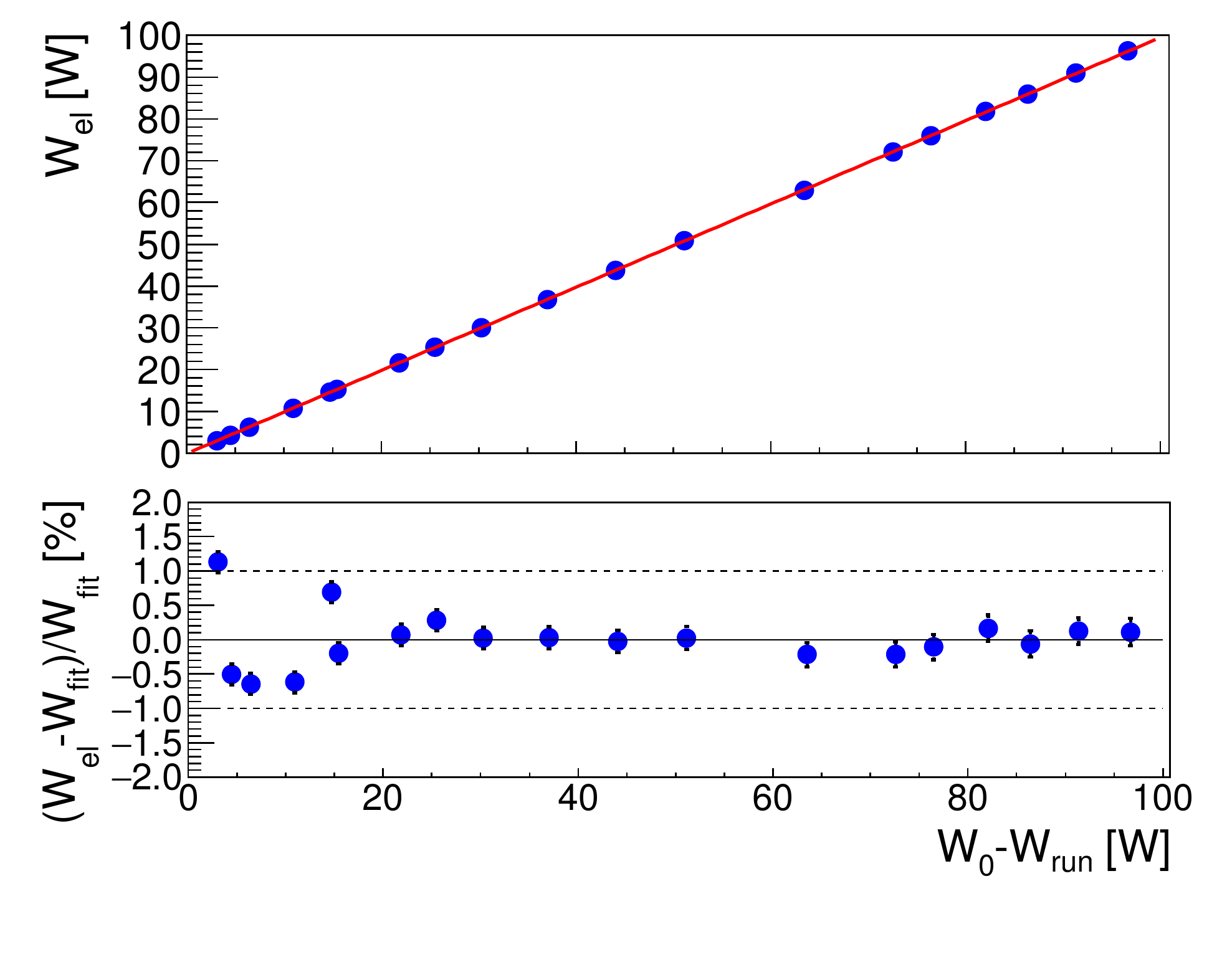}\hspace{2pc}%
\caption{Top: Calorimeter electrical calibration function obtained by fitting the electrical current reading $W_\mathrm{{el}}$ as a function of $W_{0}-W_{\mathrm{run}}$. Bottom: Relative residuals with respect to the linear fit.}
\label{fig:calocalib}
\end{figure}

Denoting with $W_0$ ($W_{\rm run}$) the power delivered by the resistors when the beam is off (on), the calorimetric power is defined as $W_{\rm cal} = W_0-W_{\rm run}$.
The calorimetric power $W_{\rm cal}$ was calibrated against the electrical power $W_{\rm el}$ measured without gas in the target chamber ($P \approx 10^{-5}$~mbar), using the calorimeter and target chamber as a Faraday cup. Results of this standard calibration procedure \cite{Ferraro18-EPJA,Cavanna14-EPJA} are displayed in fig. \ref{fig:calocalib}: the linear dependency is verified to better than 1\% over a wide power range ($W_{\rm run} = 20-100$~W).  The improved linearity compared to previous work 
\cite{Ferraro18-EPJA} was due to more accurate measurements of both the electrical and the calorimetric powers thanks to a factor of 10 reduction in residual gas pressure.
The beam current can then be determined with a 1\% uncertainty as:
\begin{equation}
    I=\frac{e\cdot W_{\rm cal}}{E_{\rm p}-\Delta E},
    \label{eq:cal}
\end{equation}
where $e$ is the elementary charge, $E_{\rm p}$ is the initial proton beam energy, and $\Delta E$ is the energy lost by the beam in the gas target (at most 3~keV), as calculated using tabulated stopping power values (with an uncertainty of 2.8\% for protons in deuterium gas) in SRIM \cite{Ziegler10-NIMB} and including the beam heating correction. 
For each run, the total number of protons $N_p$ (eq. \ref{eq:cs}) can finally be derived from the beam current as $N_{\rm p} = I \Delta t/e$ where $\Delta t$ is the live time of the run.

\section{Detection efficiency setup and measurement}
\label{sec:detector}

In the extended deuterium gas target the interaction with the proton beam can take place at different positions along the beam axis, resulting in different energies of the emitted photons (for the same beam energy) and in different geometrical angles subtended by the HPGe detector. 
Therefore, the $\gamma$-ray detection efficiency $\epsilon(z,E_\gamma)$ (eq. \ref{eq:cs}) must be carefully determined as a function of both position and energy.

For the conditions of the experiment at LUNA, the $\gamma$ rays emitted by the D(p,$\gamma$)$^3$He reaction ($Q=5.5$~MeV) have typical energies $E_\gamma = 5.5-5.8$~MeV, {\em i.e.} far away from the energy of the commonly used radioactive sources. Thus, a measurement of the detection (photo-peak) efficiency was performed using a different technique based on the well-known resonant reaction $^{14}$N(p,$\gamma_1 \gamma_2$)$^{15}$O, which produces pairs of $\gamma$ rays over a wide energy range (see sect. \ref{sec:eff}).

\begin{figure}[t!]
\includegraphics[width=19pc]{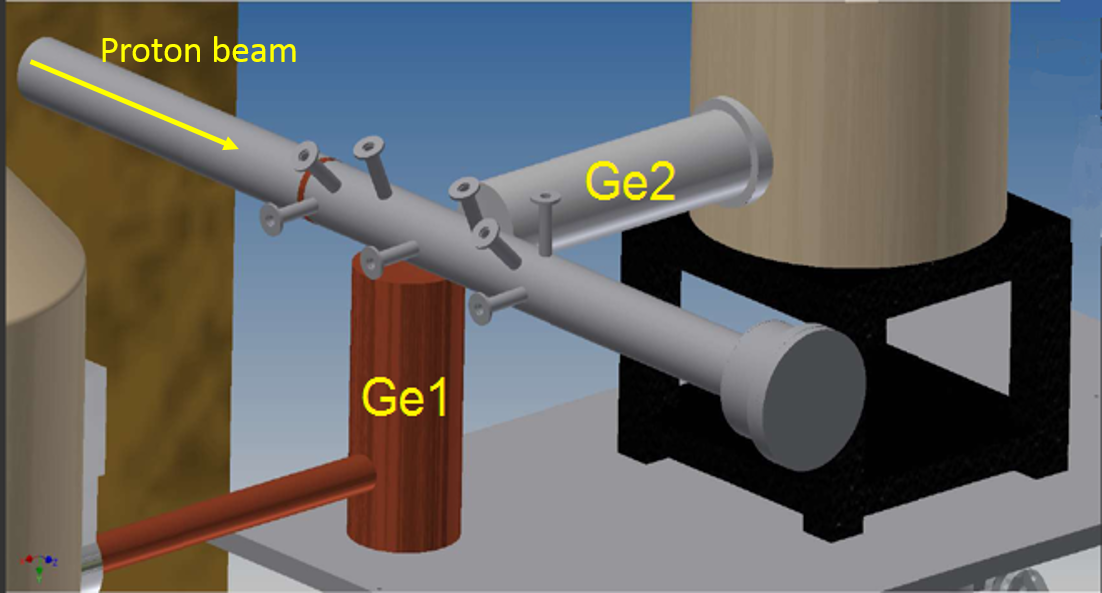}\hspace{2pc}%
\caption{3D rendering of the setup showing the two HPGe detectors used for efficiency measurements and the ports used to monitor the temperature and pressure profiles of the gas target. Errors shown are statistical only}
\label{fig:ge1ge2}
\end{figure}

For the photo-peak efficiency measurements we used the following experimental setup.
In addition to the HPGe detector (hereafter Ge1) used for the D(p,$\gamma$)$^3$He yield measurements, a second HPGe detector (hereafter Ge2) with 125\% relative efficiency was mounted on a movable platform, as shown in fig. \ref{fig:ge1ge2}, in order to change its position along the beam axis. Detector Ge2 was surrounded by a 50~mm thick lead shielding with a vertical slit 15 mm wide facing towards the reaction chamber. This lead collimator allowed us to select $\gamma$ rays generated within a well-defined position along the beam axis.

\begin{figure}[t!]
\includegraphics[width=18pc]{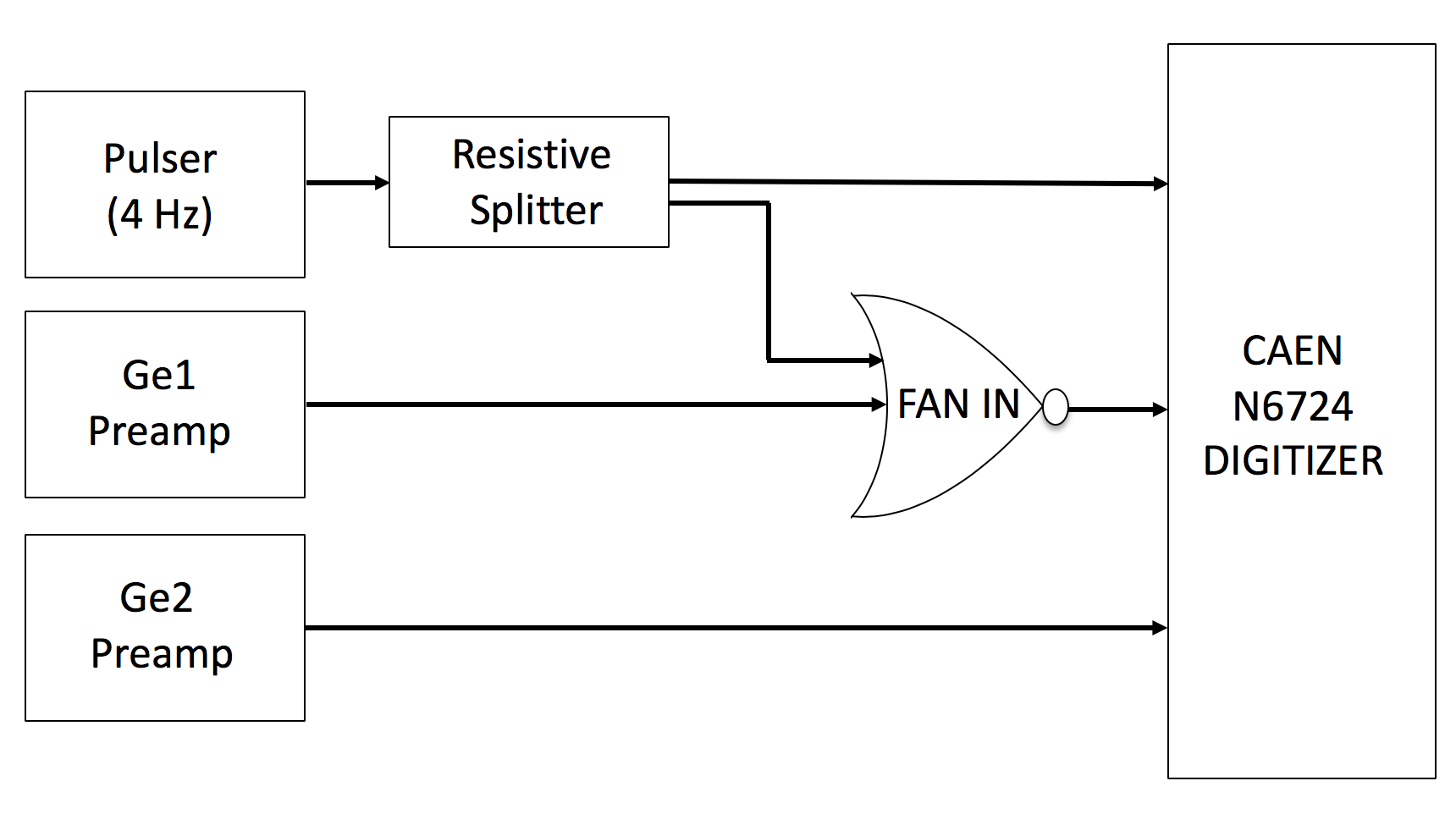}\hspace{2pc}%
\caption{Sketch of the electronic chain of the data acquisition system.}
\label{fig:DAQ}
\end{figure}

Signals from both the Ge1 and Ge2 detectors were sent to a CAEN N6724 waveform digitizer. A sketch of the electronic chain is shown in fig. \ref{fig:DAQ}. 
A pulser producing constant-amplitude signals (4 Hz) with the same shape as those produced by the Ge1 preamplifier was connected to the first channel of the CAEN digitizer. 
The same signal, together with that from the Ge1 detector preamplifier, was also used as input to a custom analog fan-in based on the THS403x amplifier whose output was fed to the second channel of the CAEN digitizer. 
Finally, the output from the Ge2 preamplifier was connected to a third channel of the CAEN module. 
A  trapezoidal filter was applied to determine the height of the signals and this information was stored, together with the signal time stamp, for offline analysis.
In this way, the DAQ dead time was quantitatively corrected for by using the pulser method \cite{Knoll10-Book}, {\em i.e.} by comparing the rate of pulser signals sent to channel 2 with the reference pulser signals sent to channel 1. 
The dead-time correction during $\mathrm{D(p,\gamma)^{3}He}$ runs was typically below 1\%. 

\subsection{Gamma-ray detection efficiency}
\label{sec:eff}

The $^{14}$N(p,$\gamma_1 \gamma_2$)$^{15}$O reaction has been studied extensively by the LUNA collaboration \cite{Formicola04-PLB, Bemmerer06-NPA}. 
At the resonant energy $E_{\rm r}=259$~keV \cite{gyurky2019} (in the centre of mass system; $\Gamma = 0.99$~keV \cite{Daigle16-PRC}) this reaction mainly proceeds through three exit channels with known branching ratios (BR), each producing two prompt $\gamma$ rays in cascade: 1) $765+6792$~keV (BR=23.0\%), 2) $1384+6172$~keV (BR=58.3\%) and 3) $2375+5181$~keV (BR=16.9\%) \cite{Marta11-PRC}. 
In order to measure the photo-peak efficiency $\epsilon(z,E_\gamma$) of Ge1 as a function of $\gamma$-ray emission position $z$ along the beam axis (with $z=0$ corresponding to the centre of the target chamber) and for all $\gamma$-ray energies above, we used the following procedure.

\begin{figure}[t!]
\includegraphics[width=20pc]{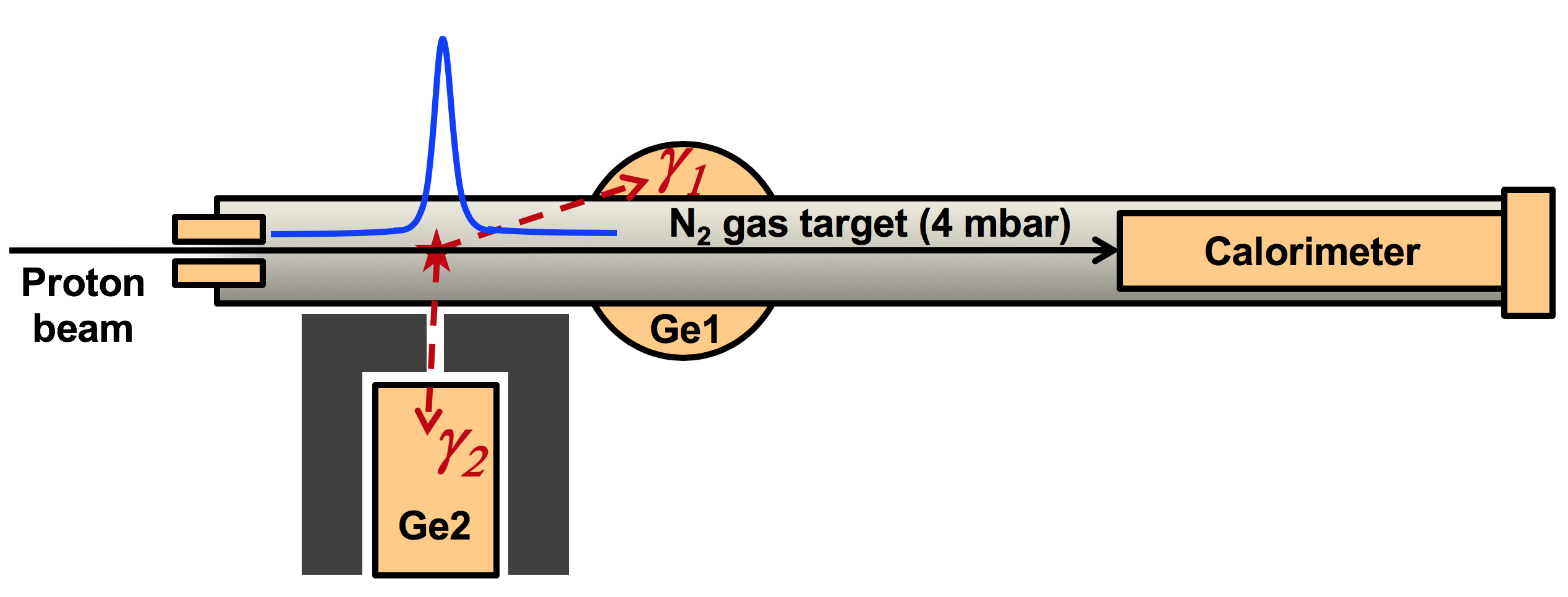}\hspace{2pc}%
\caption{Top view (not to scale) of the experimental setup
used for the efficiency measurement with the $^{14}$N(p,$\gamma_1 \gamma_2$)$^{15}$O reaction.}
\label{fig:setup2gamma}
\end{figure}

The Ge2 detector was moved to ten positions, in 30~mm steps, along the 330~mm long gas target filled with high purity N$_2$ gas at 4~mbar.
At this pressure the stopping power of the protons is about 1.3~keV/cm and the resulting spatial resonance width (FWHM) is about 0.8~cm.
By properly tuning the proton beam energy, the resonance energy $E_{\rm r}=259$~keV of the $^{14}$N(p,$\gamma_1 \gamma_2$)$^{15}$O reaction can be excited at a chosen position along the beam axis in alignment with the Ge2 detector (see fig. \ref{fig:setup2gamma}). In this way, the full energy photons detected by Ge2 mainly come from the target region facing the slit collimator, where the resonance is excited, while the off-resonance photons are suppressed by the 50~mm lead shield around the Ge2 detector. Ge2 can then be used to gate the spectrum observed on Ge1.

\begin{figure}[t!]
\includegraphics[width=22pc]{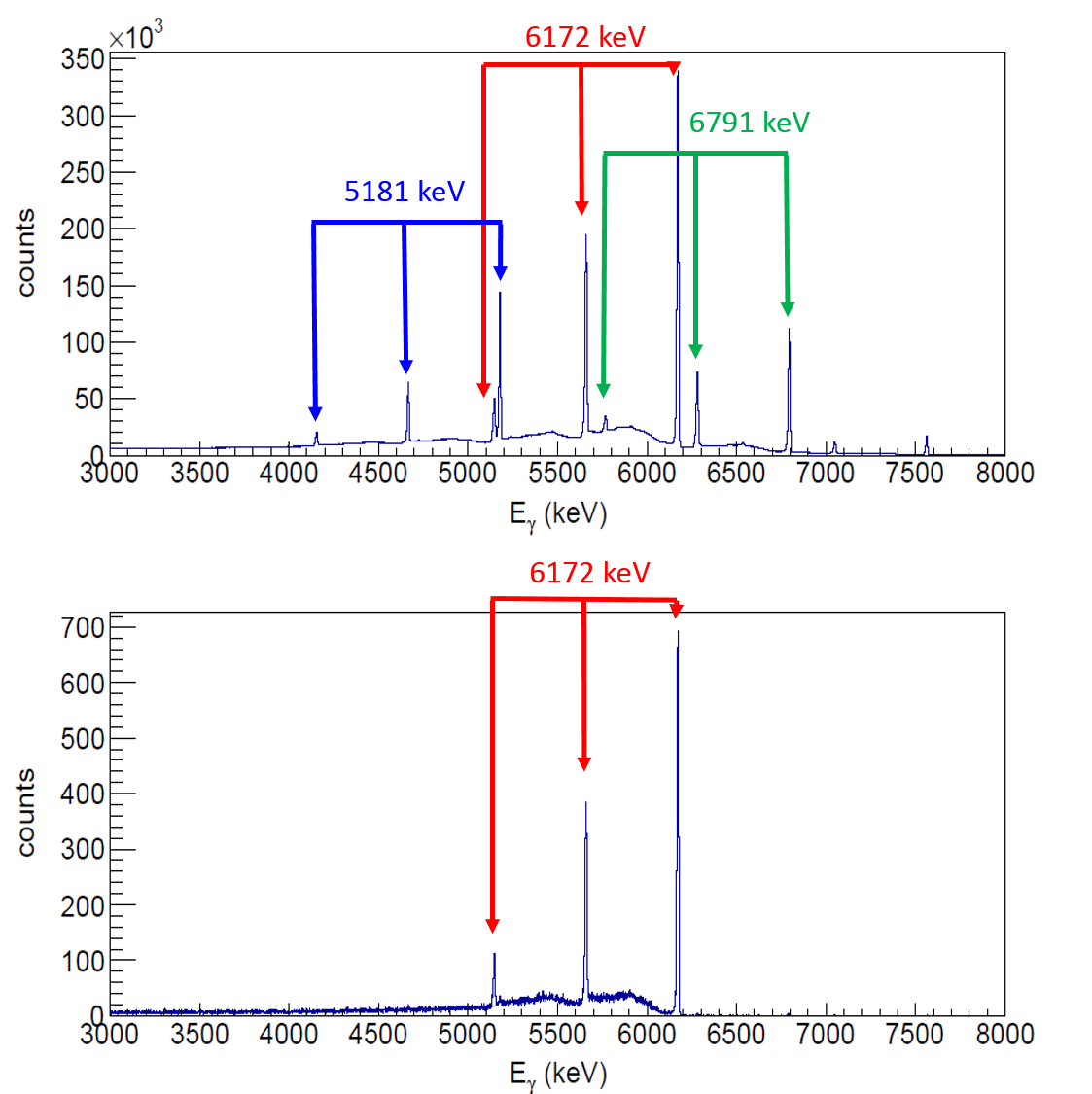}
\hspace{2pc}%
\caption{
Sample $\gamma$-ray spectrum from the $^{14}$N(p,$\gamma$)$^{15}$O reaction acquired with the Ge1 detector (top) and gated with the Ge2 detector in the $1381 < E_{\gamma_2} [\mathrm{keV}] < 1389$ energy region (bottom).
Here only the full energy peak and the single- and double-escape peaks of photons with $E_{\gamma_1} =6172$~keV can be seen.}
\label{fig:spectra}
\end{figure}

Figure \ref{fig:spectra} (top panel) shows a typical Ge1 spectrum with the full-energy, single- and double-escape peaks of the high energy photons emitted in the $^{14}$N(p,$\gamma_1 \gamma_2$)$^{15}$O reaction. Figure \ref{fig:spectra} (bottom panel) shows the spectrum obtained when imposing the condition $1381 <E_{\gamma_2} [{\rm keV}] <1389$ for the Ge2 detector. As expected, this spectrum is essentially due to photons with $E_{\gamma_1}=6172$~keV, emitted in coincidence with photons at $E_{\gamma_2} =1384$~keV detected by the Ge2 detector. 

The experimental photo-peak efficiency for Ge1 can be expressed as:
\begin{equation}
    \epsilon_{\rm data}(z,E_\gamma)=\frac{N_{\gamma_1}}{N_{\gamma_2}},
\label{eq:eff-data}
\end{equation}
where $N_{\gamma_1}$ is the net number of full energy photons detected by Ge1 (see below) when gated by Ge2 and 
$N_{\gamma_2}$ is the net number of full-energy photons detected by Ge2. The net area $N_{\gamma_2}$ was calculated using side-bands (see for example, \cite{Gilmore95}); the net area $N_{\gamma_1}$ was  corrected for  random coincidences (at most 3\%), quantified using an off-coincidence time window, as well as for dead time and instrumental effects (pileup and electronic noise) determined with the pulser method. 
Since the efficiency only depends on the  $N_{\gamma_1}/N_{\gamma_2}$ ratio, it is independent of the values (and therefore uncertainties) of beam intensity, target density, resonance strength and  branching ratios.

Finally, a set of simulations were performed to correct the measured efficiency for the angular correlation between the two $\gamma$ rays emitted in cascade and the non point-like distribution of photons detected by the Ge2 detector. The code was validated with efficiency measurements using radioactive sources ($^{60}$Co and $^{137}$Cs) and data from the $^{14}$N(p,$\gamma_1 \gamma_2$)$^{15}$O reaction.

A first set of simulations was performed using virtual point-like sources, for each $\gamma$-ray energy at each position $z$ along the beam axis. The simulations included the geometry of the detectors, according to factory drawings, and the rest of the setup. The $\gamma$ rays were tracked in the passive and active materials using the GEANT simulation toolkit.

In a second set of simulations (``full simulations''), the details of the $^{14}$N(p,$\gamma_1 \gamma_2$)$^{15}$O reaction were added, including the energy dependence of its cross section, the reaction kinematics, the angular distribution of the emitted $\gamma$ rays and the angular correlation \cite{Marta08-PRC} between pairs of photons emitted in cascade. The target density profile, thermal motion of target atoms,  beam energy losses and  angular and energy straggling \cite{Ziegler10-NIMB} were also included. Finally, the response of the detectors was taken into account by including their energy resolution as well as instrumental effects (pile-up and electronic noise) evaluated with the pulser method (see sect. \ref{sec:detector}).

\begin{figure}[t]
\includegraphics[width=22pc]{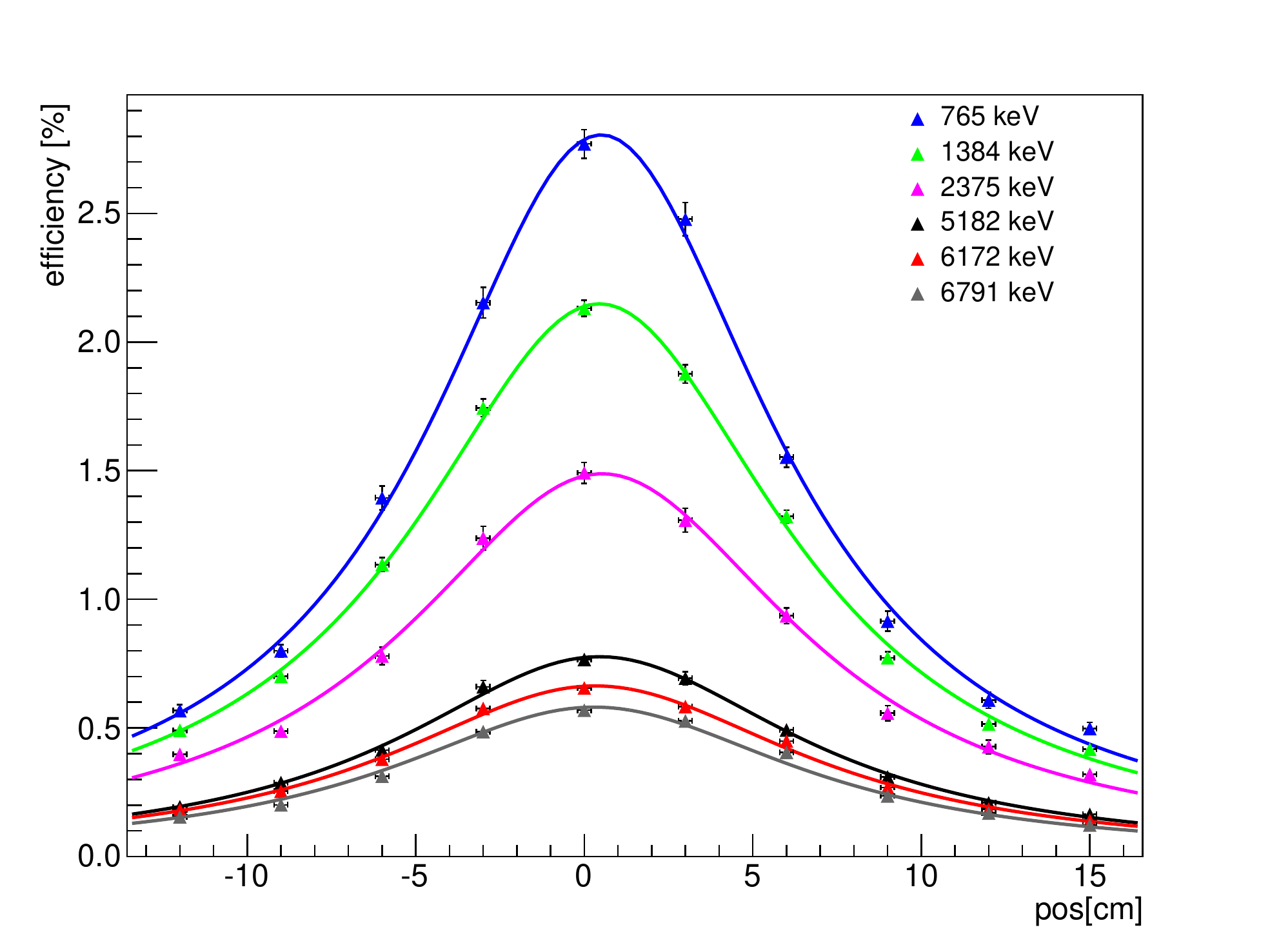}
\hspace{2pc}%
\caption{(Colour online) Photo-peak efficiency measured for six $\gamma$-ray energies as a function of source position along the beam axis ($z = 0$ corresponds to the centre of the target chamber). Errors are statistical only. Curves represent simulated efficiencies for point-like sources (see text for details).}
\label{fig:effvsz}
\end{figure}

The final experimental photo-peak efficiency ({\em i.e.}, the efficiency used in our data analysis of the D(p,$\gamma$)$^3$He cross section) can now be expressed as:
\begin{eqnarray}
{\epsilon(z,E_\gamma)=
\epsilon_{\rm data}(z,E_\gamma)
\frac{\epsilon_{\rm pl}(z,E_\gamma)}
{\epsilon_{\rm full}(z,E_\gamma)}},
\label{eq:eff}
\end{eqnarray}
in terms of the measured efficiency $\epsilon_{\rm data}$ (eq. \ref{eq:eff-data}), corrected by the ratio of the simulated efficiencies for a point-like source $\epsilon_{\rm pl}$, and for an extended source $\epsilon_{\rm full}$ (from the full simulation), respectively.
The ratio between these two efficiencies largely cancels out possible differences between experimental and simulated setup.
A comparison between the final experimental and simulated efficiencies, $\epsilon(z,E_\gamma)$ and $\epsilon_{\rm pl}(z,E_\gamma)$, as a function of position is shown in fig. \ref{fig:effvsz} for the six $\gamma$-ray energies of the $^{14}$N(p,$\gamma_1 \gamma_2$)$^{15}$O reaction. 

The average discrepancy (obtained by averaging over all measured positions, each weigh\-ted for the solid angle subtended by the detector at that position) between $\epsilon(z,E_\gamma)$ and $\epsilon_{\rm pl}(z,E_\gamma)$ turned out to be 2\% at most, and this value was taken as the error associated to the Ge1 $\gamma$-ray photo-peak efficiency for the D(p,$\gamma$)$^3$He reaction study.

\section{Angular distribution considerations}
\label{sec:PSA}

\begin{figure}[t!]
\includegraphics[width=22pc]{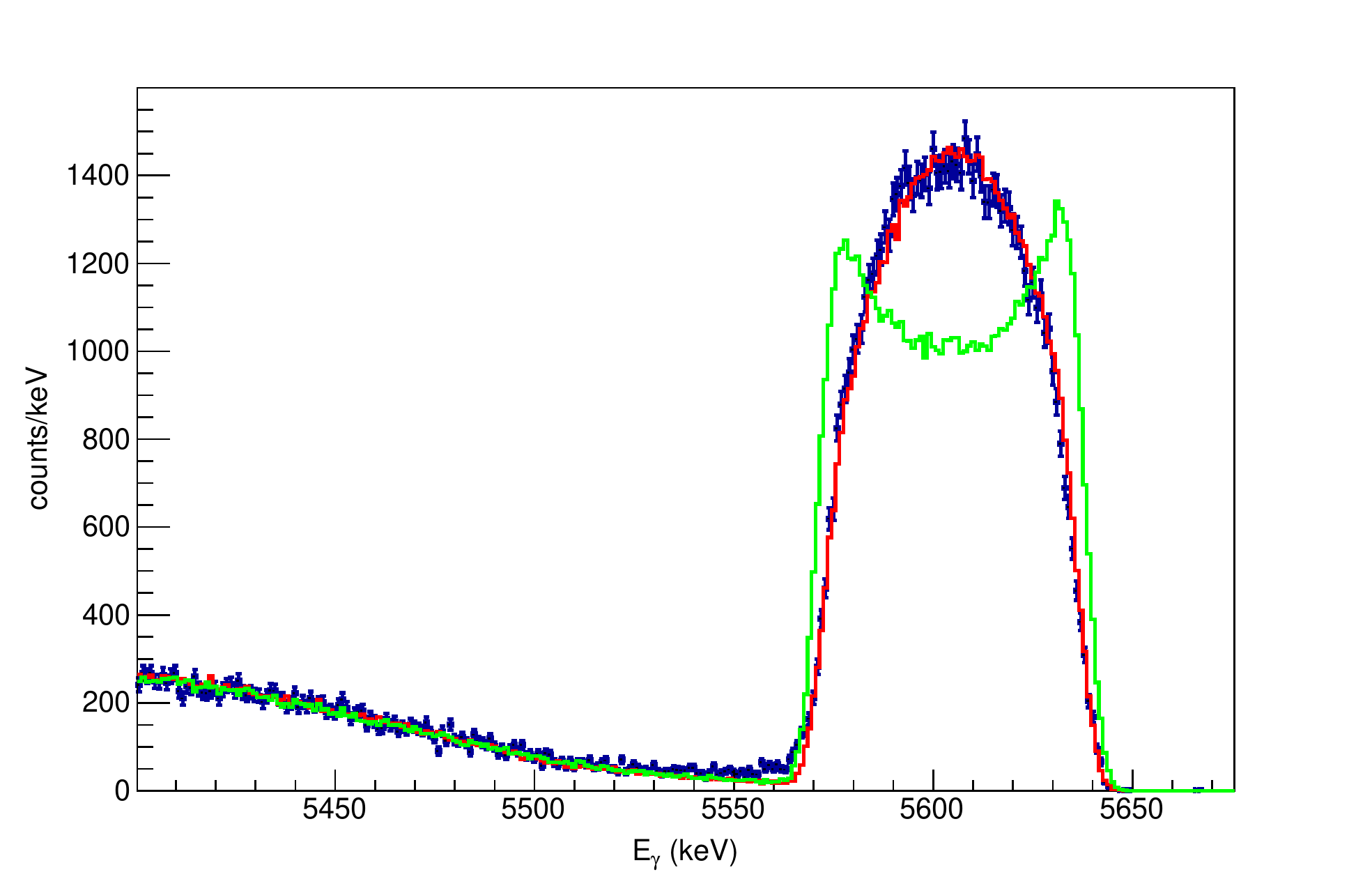}
\hspace{2pc}%
\caption{(Colour online) Experimental (blue points) and simulated spectra for the $\mathrm{D(p,\gamma)^{3}He}$ reaction, assuming isotropic (green) and \emph{ab initio} (red) angular distributions \cite{Marcucci18-private} at $E_{\mathrm{p}}=175$ keV.}
\label{fig:mcdata}
\end{figure}

The final ingredient entering the cross section in eq. (\ref{eq:cs}) concerns the term $W(z)$ accounting for the angular distribution of the $\gamma$ rays emitted by the D(p,$\gamma$)$^3$He reaction. 
The impact of $W(z)$ on the total error budget was evaluated by Monte Carlo simulations assuming both isotropic and {\em ab initio} \cite{Marcucci18-private} distributions. 
Specifically, the energy $E_{\gamma}$ of the emitted photon depends on its polar angle $\theta_{\rm lab}$ and on the beam energy $E_p$ according to the expression (with $c=1$):
$${E_\gamma = \frac{m_{\mathrm{p}}^2+m_{\mathrm{D}}^2-m_{\mathrm{He}}^2+2(E_{\mathrm{p}}+m_{\mathrm{p}})m_{\mathrm{D}}}{2(E_{\mathrm{p}}+m_{\mathrm{p}}+m_{\mathrm{D}}-p_{\mathrm{p}}\cos \theta_{\mathrm{lab}})}},
$$
where $m_{\rm p}$, $m_{\rm D}$ and $m_{\rm He}$ are the masses of the nuclides involved in the reaction and $p_{\rm p} = \sqrt{E_{\rm p} \left(E_{\rm p} + 2m_{\rm p}\right)}$ is the proton momentum.
For the experimental setup used (sect. \ref{sec:setup}) the angular acceptance of the Ge1 detector was 
$\theta_{\rm lab} \simeq 15^{\circ}-165^{\circ}$, corresponding to a range of $\gamma$-ray energies $E_\gamma \simeq 5.7-5.8$~MeV at $E_{\rm p}=390$~keV.
As a result, the full energy peak is broadened by kinematics, while its shape depends on the photon angular distribution. 
Figure \ref{fig:mcdata} shows a comparison between experimental data (blue points) and the D(p,$\gamma$)$^3$He simulated spectra at $E_{\rm p}=175$~keV assuming an isotropic (green curve) or an \emph{ab initio} (red curve) distribution \cite{Marcucci18-private} of the emitted photons. 
Despite the different shapes, the net areas of full energy peaks in the two cases (isotropic and \emph{ab initio} distributions) only differ by about $2\%$ confirming that with our setup the integrated number of counts depends weakly on the photon angular distribution. 
To evaluate the systematic uncertainty associated with angular distribution effects, we varied the coefficient $a_2$ of the Legendre polynomial in the Monte Carlo simulation  distribution between values -1 and -0.5. 
We observed an overall discrepancy of $\pm$0.5\% at most in the net areas of the full energy peaks. 
The same procedure was repeated at the extremes of our energy range, i.e. at $E_{\rm p}=50$~keV and $E_{\rm p}=400$~keV, obtaining consistent results. 
As a further check we used the Legendre expansion coefficients reported by Schmid \emph{et al.} \cite{Schmid97-PRC} at 50\,keV bombarding energy as inputs to our simulation. 
The photo-peak areas obtained with the two distributions (Schmid and {\em ab initio}) differ by less than 0.03\%.
Assuming either Schmid or Marcucci angular distributions, the fraction of $\gamma$ rays falling outside the angular range covered by our setup is about 0.7\% and is properly accounted for in eq. (\ref{eq:cs}).
The systematic error arising from angular distribution effects was assigned to be $\pm$0.5\%, as discussed above.

\section{Summary of systematic uncertainties}
\label{sec:errors}

The sources of systematic errors affecting the D(p,$\gamma$)$^{3}$He cross-section evaluation, together with the methods used to quantify them, are listed in table \ref{tab:errors}.
The 2\% uncertainty on the $\gamma$-ray detection efficiency (sect. \ref{sec:eff}) dominates the total error budget at all beam energies investigated. 
The total error on the target density (1.1\%) is determined by summing in quadrature the uncertainties in temperature and pressure profile (1\%) and in the beam heating effect (0.5\%). 
The uncertainties on beam current and angular distribution are respectively 1\% (sect. \ref{sec:current}) and 0.5\% (sect. \ref{sec:PSA}).
All other sources of systematic errors, listed in table \ref{tab:errors}, remain negligible. We note that the statistical error on the number of detected $\gamma$ rays $N_\gamma$ (eq. \ref{eq:cs}), including background subtraction, was typically below 1\%. 
The overall systematic error achieved over the energy range $E_{\rm p} = 50-400$~keV covered for the  $\mathrm{D(p,\gamma)^3He}$ reaction study at LUNA remains below 3\%.
\begin{table}[t!]
\centering
\caption{Contributions to the overall systematic uncertainty in the in the $\mathrm{D(p,\gamma)^{3}He}$ S factor arising from different sources. Values  shown refer to a representative energy $E_{\rm p} = 200$~keV ($E_{\rm cm} = 133$~keV).}
\hspace{1pt}
\begin{tabular}{|l|l|l|} \hline
Source  & Method & $\Delta$S/S\\
\hline
Beam energy  & Direct measurement &  $ 0.2\%$
\\
Energy loss   & Low gas pressure &  $ 0.04\%$
\\
T and P profiles & Direct measurement &  $ 1.0\%$                  \\
Beam heating    & Direct measurement &  $ 0.5\%$                  \\
Gas purity & Data sheet &  $  0.1\% $                  \\
Beam current & Calorimeter calibration &  $  1.0\%$                   \\
Efficiency & Direct measurement &  $  2.0\%$                  \\
Instrumental effects & Pulser method &  $  0.2\% $                   \\
Angular distribution & Simulations &  $  0.5\%$                    \\
\hline
Total & & $2.6\%$ \\
\hline
\end{tabular}
\label{tab:errors}
\end{table}
This result represents a significant improvement with respect to the $10-20$\% systematic uncertainties affecting previous data at energies most relevant to BBN \cite{Ma97-PRC,Tisma19-EPJA}. 
The cross-section results obtained at LUNA and their implications in cosmology and particle physics will be published in a forthcoming paper.

\section{Conclusions}
\label{sec:summa}

Direct observations of deuterium abundance can be used to tightly constrain the universal baryon density and the number of relativistic particles existing in the early Universe as long as accurate BBN predictions are provided.  
In this context, precise nuclear reaction rates are crucial for each of the relevant reactions in the BBN network. 
Among the reactions that affect the primordial deuterium abundance, the $\mathrm{D(p,\gamma)^3He}$ remains the least well-known. 
A significant effort was devoted by the LUNA collaboration towards a renewed measurement of its cross section with unprecedented precision.
Here, we reported on a series of commissioning measurements leading to an overall systematic uncertainty below 3\%.
This accuracy will enable theoretical calculations of deuterium abundance to the same level as that observed.

\section*{Acknowledgments}
The authors acknowledge the invaluable contribution of Donatello Ciccotti for his support during all phases of the experiment at LUNA, Marco D'Incecco for his work on custom electronics, Massimiliano De Deo for the implementation of the data acquisition system, and Giuliano Sobrero for the development of the new gas target control panel. We also acknowledge the mechanical workshop at LNGS, INFN sez. Bari and Dipartimento Interateneo di Fisica Bari. 
This work has mainly been supported by INFN, with contributions by Helmholtz Association (ERC-RA-0016), DFG (BE4100/4-1), NKFIH (K120666), COST (ChETEC CA16117),  STFC-UK and the University of Naples Compagnia di San Paolo (STAR).

\bibliographystyle{plainnat}

\bibliography{Mossa}
\end{document}